\documentclass[preprint,12pt,authoryear]{elsarticle}

\usepackage{amssymb}
\usepackage{amsmath}

\usepackage{epstopdf, epsfig}
\usepackage{amstext}
\usepackage{amsfonts,amssymb}
\usepackage{algorithm}
\usepackage{algorithmicx}
\usepackage{bm}
\usepackage{graphicx}
\usepackage{subcaption}
\usepackage{color}
\usepackage{subcaption}
\usepackage[HTML]{xcolor}

\definecolor{steelcyan}{RGB}{70,130,180}
\definecolor{mygray}{RGB}{89,89,89}
\definecolor{mypink}{RGB}{255,191,204}
\definecolor{mycyan}{RGB}{51,179,230}
\definecolor{EnkfColor}{HTML}{1f77b4}
\definecolor{EnKFMLCcolor}{HTML}{ff7f0e}
\newcommand{\colorSquare}[1]{{\color{#1}\rule{1.6em}{0.6em}}}

\usepackage{booktabs} 

\DeclareMathAlphabet{\pazocal}{OMS}{zplm}{m}{n}

    \def\dashL{\bm{\mbox{--~--~--}}}

\journal{Computer Methods in Applied Mechanics and Engineering}

\begin{document}

\begin{frontmatter}



\title{Enabling High-Accuracy Data Assimilation with Limited Ensembles via Machine Learning-Based Covariance Correction} 


\author[1]{Z. Yao\corref{coa1}} 
\author[1]{Z. Li\corref{coa1}}
\author[2]{L. Zhao}
\author[3]{Z. Liu}
\author[4]{Z. Lu}
\author[5]{S. Kim}
\author[1,6]{G. Wang\corref{cor1}}

\affiliation[1]{organization={Centre for Regional Oceans, Department of Ocean Science and Technology, and State Key Laboratory of Internet of Things for Smart City, University of Macau},
            city={Macau}}
\affiliation[2]{organization={Macau Millennium College},
            city={Macau}}
\affiliation[3]{organization={School of Naval Architecture and Ocean Engineering, Huazhong University of Science and Technology},
            city={Wuhan},
            state={Hubei},
            country={China}}
\affiliation[4]{organization={Ningbo Institute of Dalian University of Technology},
            city={Ningbo},
            state={Zhejiang},
            country={China}}
\affiliation[5]{organization={Department of Naval Architecture and Ocean Engineering, Hongik University},
            city={Sejong},
            country={Republic of Korea}}
\affiliation[6]{organization={Zhuhai UM Science and Technology Research Institute},
            city={Zhuhai},
            state={Guangdong},
            country={China}}

\cortext[coa1]{The authors contribute equally to this paper.}
\cortext[cor1]{Corresponding author.}
\ead{wanggy@um.edu.mo}

\begin{abstract}
Data assimilation (DA) integrates numerical model forecasts with observations to achieve the optimal state estimation. Ensemble-based methods, such as the ensemble Kalman filter (EnKF), are widely used for state estimation for high-dimensional and nonlinear dynamic systems. However, their performance strongly depends on the ensemble size, therefore causing a tradeoff problem between analysis accuracy and computational cost. To address this problem, this study presents a machine learning-based EnKF framework that maintains high accuracy with a relatively small ensemble size. Specifically, a multilayer perceptron (MLP) function is built to predict the difference between the forecast error covariances estimated from a limited ensemble and a sufficiently large ensemble, with the latter being assumed to be an accurate approximation of the underlying truth. This predicted covariance difference term is then incorporated into the EnKF algorithm via an element-wise scaling strategy, resulting in an amended forecast covariance matrix that better approximates the true uncertainty level and sequentially produces more accurate analysis results. To demonstrate the feasibility and robustness of the proposed algorithm, we perform a set of numerical experiments with the Lorenz-63 and Lorenz-96 systems under various configurations, and the results consistently indicate that the proposed algorithm can significantly outperform the standard EnKF with the same limited ensemble size, by achieving notably higher analysis accuracy while remaining computationally efficient. This approach provides a practical and feasible pathway to accurate and computationally efficient data assimilation for high-dimensional and nonlinear dynamic systems.
\end{abstract}







\end{frontmatter}



\section{Introduction}
\label{sec:Intro}
Data assimilation (DA) is a mathematical framework that incorporates the observations into numerical models to obtain an optimal state estimation via Bayesian inference~\citep{pandya2022review}. Over the past few decades, DA methods have proven to be indispensable in diverse fields, including but not limited to weather forecasting~\citep{navon2009data,rabier2005overview}, physical oceanography \citep{anderson1996data,martin2015status,martin2025data,wang2021phase,wang2022phase}, turbulence modeling~\citep{duraisamy2019turbulence,zhang2022ensemble}, and material modeling~\citep{jin2026ensemble,matsuzaki2018estimation}. Among the various DA approaches, ensemble-based methods, particularly the ensemble Kalman filter (EnKF)~\citep{evensen1994sequential,evensen2003ensemble}, have gained widespread popularity due to their inherent natural ability to handle nonlinear systems and ``plug-and-play'' architecture. 

However, ensemble-based DA methods often face a practical and challenging dilemma, i.e., better accuracy usually comes at the price of higher computational cost \citep{petrie2010ensemble,whitaker2012evaluating}. Specifically, accurate state estimation usually relies on a sufficiently large ensemble to properly characterize the true but usually poorly known forecast error covariances, i.e., uncertainty level; and a limited ensemble size may cause the rank-deficiency issues, which are usually manifested as spurious correlations and underestimation. However, the computational cost scales nearly linearly with the ensemble size, making large-ensemble DA impractical for high-dimensional dynamic systems. Two ad hoc remedies, including localization \citep{buehner2007spectral} and inflation \citep{kang2012estimation}, have been proposed and widely applied to address these issues. However, both methods introduce additional parameters that often require careful and time-consuming tuning~\citep{choi2025sampling}, which therefore cannot effectively balance the analysis accuracy and computational efficiency.




In recent years, machine learning (ML) has emerged as a promising tool to address the challenges faced by ensemble-based DA methods. Current efforts can be generally categorized into three types. The first type focuses on using neural networks (NNs) as surrogate models to accelerate the forecast step, i.e., replacing traditional computationally expensive models with learned functions~\citep{brajard2020combining,sun2025online}. Although computationally efficient, such surrogates often struggle with generalization to out-of-distribution scenarios. The second category applies ML to enhance the analysis step, mainly by learning the update formula to handle missing or low-quality data~\citep{arcucci2021deep,wu2021fast}. However, this type of methods may experience performance degradation in the scenario of low DA frequency, as shown by~\citet{li2025small}. The third type, which is also the one most relevant to uncertainty quantification (UQ), employs ML to improve the representation of forecast uncertainty, either by ameliorating the forecast ensemble members or correcting deficiencies of the preliminary covariance estimations. This direction is especially meaningful and promising for operational DA, where the uncertainty level information is essential. For instance, ~\citet{irrgang2020machine} and~\citet{gronquist2021deep} demonstrate that NNs can be utilized to predict ensemble spreads or spatiotemporal uncertainty maps with deterministic inputs, thereby significantly reducing the computational cost of running full ensemble systems. Similarly, 
\citet{sacco2022evaluation} evaluate the capacity of different NNs to identify uncertainty sources, such as imperfect initial conditions and model parameters, using both direct (ensemble-based) and indirect (analysis-based) training strategies. Building on this,~\citet{sacco2024line} propose one algorithm, which leverages convolutional neural networks to realize the online estimation of forecast error covariances from a single model integration for sequential DA. Despite these advances, two limitations still remain outstanding. First, these methods usually rely on the Gaussian assumption for error distributions, which may not always hold for highly nonlinear or chaotic systems and can lead to mis-calibrated uncertainty estimates. Second, most of the existing methods are not designed to handle low-frequency DA scenarios, where the forecast error covariance can vary significantly between assimilation cycles. This limitation can reduce their reliability in practical application situations with temporally sparse observation updates.

In this study, we propose a novel machine learning-based covariance correction framework for the ensemble Kalman filter. In contrast to the existing approaches that get rid of the ensemble completely, our method retains a relatively small ensemble while employing a multilayer perceptron (MLP) to predict a correction term, which is then used to refine the preliminary estimated covariances. By incorporating this ML-predicted correction into the standard EnKF workflow via an element-wise scaling strategy, we aim to maintain the flow-dependent nature of the uncertainty while significantly mitigating the errors induced by the limited ensemble size. We demonstrate the feasibility and  robustness of this approach through a set of numerical experiments based on Lorenz-63 and Lorenz-96 systems under various configurations, showing that it can significantly improve the overall accuracy while preserving computational efficiency.

The paper is organized as follows. $\S$~\ref{sec:Methods} presents the methodological framework, including the EnKF formulation, the ML-based covariance correction strategy, and the complete assimilation algorithm. $\S$~\ref{sec:Numerical_Experiments} describes the experimental setup and presents the results of Lorenz-63 and Lorenz-96 systems, as well as the comprehensive sensitivity analyses. Finally, $\S$~\ref{sec:Conclusion} briefly summarizes this work.

\section{Methods}
\label{sec:Methods}

\subsection{Overview of the Proposed Framework}
\label{sec:overview}

\begin{figure}
    \centering
    \includegraphics[width=1\linewidth]{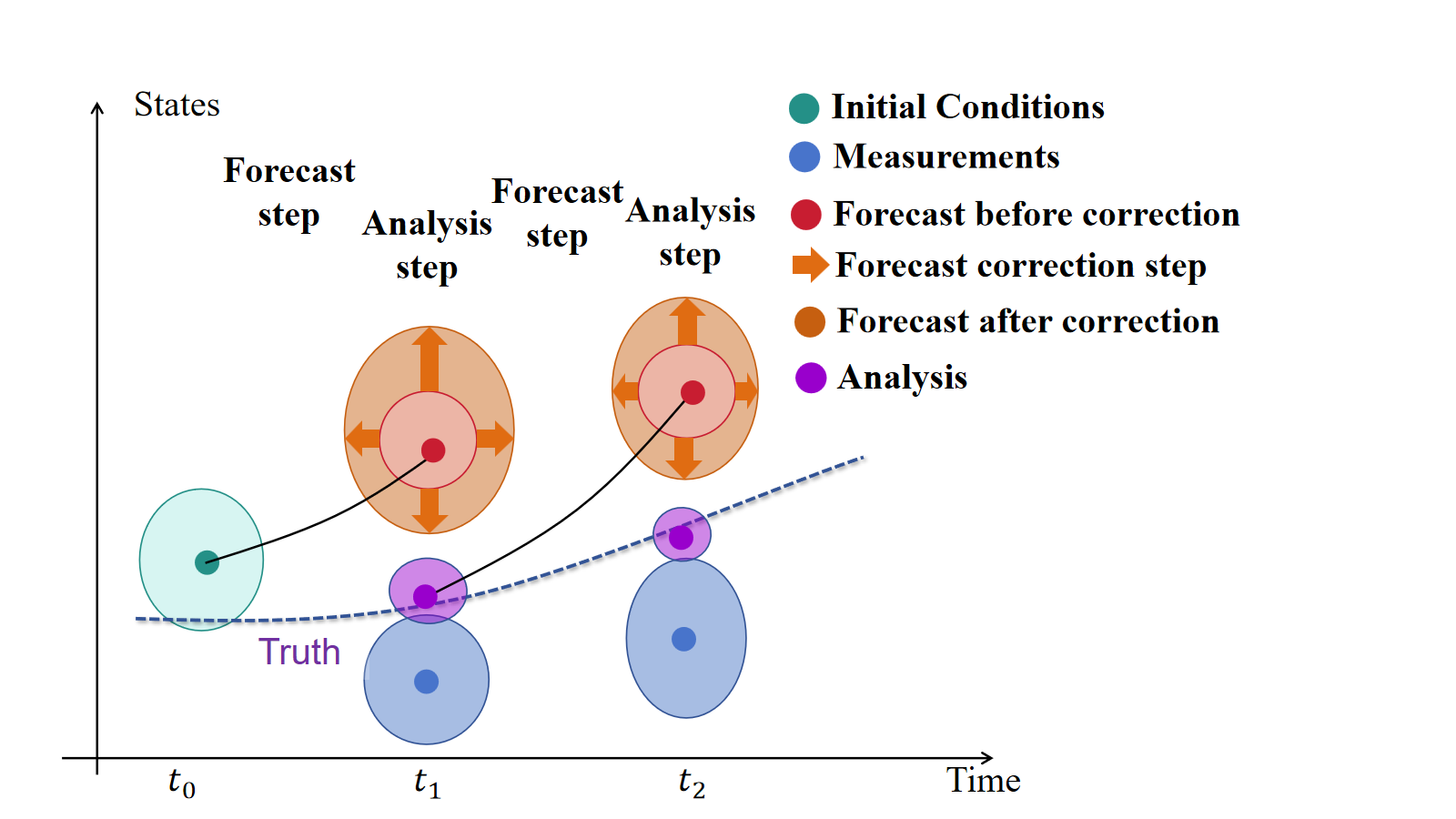}
    \caption{Schematic illustration of the proposed algorithm for correcting forecast covariance. For the purpose of brevity, only the ensemble mean is shown, omitting individual members.}
    \label{fig:ENKFFCNN}
\end{figure}

The proposed data assimilation framework, which integrates the EnKF with a ML-based covariance correction function (named as EnKF-MLC), is illustrated in Fig.~\ref{fig:ENKFFCNN}. The simulation is initialized with a relatively small ensemble (size $\mathfrak{N}$) of initial conditions, which are randomly sampled from a presumed distribution. Then the forecast step is performed for each ensemble member by running the forward integration model $\mathcal{M}$, which results in the forecast ensemble at the first DA time instant $t=t_1$. However, the forecast error covariance matrix ($\boldsymbol{P}^\mathfrak{N}_{f,1}$) estimated from this small ensemble may be inaccurate, exhibiting either underestimation or spurious correlations and potentially leading to suboptimal analysis or even filter divergence. To address this issue, a ML-based function is applied to perform the element-wise correction on $\boldsymbol{P}^\mathfrak{N}_{f,1}$, with the forecast ensemble members amended simultaneously (see details in $\S$\ref{sec:ml_correction}). Then, the analysis step is performed by implementing the standard EnKF (see details in $\S$\ref{sec:enkf}), while using the corrected forecast ensemble and covariance. Afterwards, the forecast, ML-based correction, and analysis steps are performed sequentially for all future DA time windows $\left[t_j,~t_{j+1}\right],~j~\geq~1$. 

\subsection{ML-based forecast correction}
\label{sec:ml_correction}

At one DA time instant $t_j$, we assume that the preliminary (uncorrected) forecast ensemble with the size of $\mathfrak{N}$ has been obtained as
\begin{equation}
\boldsymbol{S}_{f, j}^{\mathfrak{N}}=\left[\boldsymbol{s}^{(1)}_{f, j},\boldsymbol{s}^{(2)}_{f, j},\dots \boldsymbol{s}^{(n)}_{f, j}, \dots \boldsymbol{s}^{(\mathfrak{N}-1)}_{f, j},\boldsymbol{s}^{(\mathfrak{N})}_{f, j}\right],
\end{equation}
and the corresponding forecast error covariance matrix $\boldsymbol{P}_{f,j}^{\mathfrak{N}}$ is calculated as
\begin{equation}
    \boldsymbol{P}_{f,j}^{\mathfrak{N}}=\frac{1}{\mathfrak{N}-1}\Sigma_{n=1}^\mathfrak{N}(\boldsymbol{s}^{(n)}_{f,j}-\bar{\boldsymbol{s}}_{f,j})(\boldsymbol{s}^{(n)}_{f,j}-\bar{\boldsymbol{s}}_{f,j})^T,
\label{eq:covf}
\end{equation}
where $\bar{\boldsymbol{s}}_{f,j}=(1/\mathfrak{N})\sum_{n=1}^\mathfrak{N}\left(\boldsymbol{s}^{(n)}_{f,j}\right)$ is the forecast mean. Our goal is to build a ML-based function that can produce a correction term $\Delta\boldsymbol{P}_{f,j}$, such that $\boldsymbol{P}_{f,j}^{\mathfrak{N}} + \Delta\boldsymbol{P}_{f,j}$ can better approximate the true forecast error covariance $\boldsymbol{P}_{T,j}$. Since $\boldsymbol{P}_{T,j}$ is usually unknown in practice, we utilize the covariance matrix obtained with a relatively large ensemble size $\mathcal{N}$, denoted as $\boldsymbol{P}_{f,j}^{\mathcal{N}}$, as a surrogate. Therefore, the covariance correction term is expressed as
\begin{equation}
    \Delta\boldsymbol{P}_{f,j}=\boldsymbol{P}^{\mathcal{N}}_{f,j}-\boldsymbol{P}^{\mathfrak{N}}_{f,j}.
    \label{eq:correct}
\end{equation}
To predict $\Delta\boldsymbol{P}_{f,j}$, a MLP function is constructed as
\begin{equation}
\Delta\boldsymbol{P}_{f,j}=\mathcal{F}\left(\boldsymbol{P}_{f,j}^{\mathfrak{N}}, \boldsymbol{P}_{f,j-1/T_{\mathrm{DA}}}; \boldsymbol{\theta}\right),
\label{eq:correction}
\end{equation}
where $\boldsymbol{\theta}$ represents the trainable parameters. $\boldsymbol{P}_{f,j-1/T_{\mathrm{DA}}}$ is the (optimal) error covariance matrix of the immediately preceding numerical model time instant, with $T_{\mathrm{DA}}$ being the DA interval. Specifically, if $T_{\mathrm{DA}}$ includes only one time step ($\Delta t$) of the numerical model (i.e., $T_{\mathrm{DA}}=\Delta t$), $\boldsymbol{P}_{f,j-1/T_{\mathrm{DA}}}$ is taken as the analysis covariance matrix of the last DA time instant $\boldsymbol{P}_{a,j-1}$, which can be obtained after implementing EnKF with corrected forecast at $t=t_{j-1}$. However, if $T_{\mathrm{DA}}$ includes multiple numerical model time steps, we need to first initialize the current DA time window with the latest analysis ensemble $\boldsymbol{S}_{a,j-1}^\mathfrak{N}$ (which is obtained with the corrected forecast), run $\mathcal{M}$ until $t=t_{j}-\Delta t$, and finally estimate $\boldsymbol{P}_{f,j-1/T_{\mathrm{DA}}}$. Therefore, for both circumstances of $T_{\mathrm{DA}}$, the accumulative impact of all past corrections is incorporated, which is consistent with the flow-dependent nature of EnKF. Also it should be noted that, for $t=t_1$, $\boldsymbol{P}_{f,j-1/T_{\mathrm{DA}}}$ is unavailable.  In this regard, we directly utilize the covariance given by the large ensemble $\boldsymbol{P}_{f,j-1/T_{\mathrm{DA}}}^{\mathcal{N}}$ (produced from a separate run). Therefore, when applying this method, it is necessary to run the standard EnKF with the large ensemble for the first DA window. For the succeeding DA windows, we strictly follow the method described above. Once the correction term $\Delta\boldsymbol{P}_{f,j}$ is predicted, the forecast error covariance matrix is updated as
\begin{equation}{\boldsymbol{P}}^{\mathfrak{N}}_{f,j}\Leftarrow
     {\boldsymbol{P}}^{\mathfrak{N}}_{f,j} + \Delta\boldsymbol{P}_{f,j}.
\label{eq:pcorrected}
\end{equation}

Last but not least, to ensure the consistency between the ensemble members and the covariance matrix, the forecast ensemble is then resampled from a multivariate normal distribution with the mean $\bar{\boldsymbol{S}}^{\mathfrak{N}}_{f,j}$ and corrected covariance ${\boldsymbol{P}}^{\mathfrak{N}}_{f,j}$.

\subsection{Ensemble Kalman Filter}
\label{sec:enkf}

The proposed EnKF-MLC framework utilizes the stochastic version of EnKF \citep{burgers1998analysis,van2020consistent}. Specifically, at any DA time instant, with a general ensemble size $N$ (i.e., $N$ can be either $\mathfrak{N}$ or $\mathcal{N}$), we start with the forecast ensemble
\begin{equation}
\boldsymbol{S}^N_{f}=\left[\boldsymbol{s}^{(1)}_{f},\boldsymbol{s}^{(2)}_{f},\dots \boldsymbol{s}^{(n)}_{f}, \dots \boldsymbol{s}^{(N-1)}_{f},\boldsymbol{s}^{(N)}_{f}\right],
\end{equation}
and calculate $\boldsymbol{P}_{f}^{{N}}$ by following Eq.~\eqref{eq:covf}. Here we have dropped off the subscript denoting time for the purpose of conciseness.  Afterwards, the corresponding measurement ensemble is generated as 
\begin{equation}
     \boldsymbol{S}^{N}_{m}=\left[\boldsymbol{s}^{(1)}_{m},\boldsymbol{s}^{(2)}_{m},\dots \boldsymbol{s}^{(n)}_{m}, \dots \boldsymbol{s}^{(N-1)}_{m},\boldsymbol{s}^{(N)}_{m}\right],
\end{equation}

\begin{equation}
\boldsymbol{s}^{(n)}_{m}=\bar{\boldsymbol{s}}_{m}+\boldsymbol{\eta}^{(n)},
\end{equation}
where $\bar{\boldsymbol{s}}_{m}$ is the measurement mean and $\boldsymbol{\eta}^{(n)}$ is the measurement perturbation term, which is usually sampled from a presumed distribution with zero-mean and covariance $\boldsymbol{R}$. Then the DA operation is implemented as

\begin{equation}
     \boldsymbol{S}^{N}_{a} = \boldsymbol{S}^{N}_{f} + \boldsymbol{K} \left(\boldsymbol{S}_m^{N} - \boldsymbol{H} \boldsymbol{S}_{f}^{N} \right),
\label{eq:enkf}
\end{equation}
where
\begin{equation}
     \boldsymbol{K} = {\boldsymbol{P}}^{N}_{f} \boldsymbol{H}^T \left( \boldsymbol{H} {\boldsymbol{P}}^{N}_{f} \boldsymbol{H}^T + \boldsymbol{R} \right)^{-1},
\label{eq:Kalman gain}
\end{equation}
is the Kalman gain. In this study, we only consider the linear  mapping operator $\boldsymbol{H}$, while the proposed method can be easily extended to the non-linear case. To implement the proposed framework, we just need to set $N=\mathfrak{N}$, and replace $\boldsymbol{P}_f^{N}$ and $\boldsymbol{S}_f^{N}$ with their corrected counterparts from $\S$~\ref{sec:ml_correction}.\\
\vspace{0.3cm}

The complete workflow of the proposed EnKF-MLC framework is summarized in Algorithm~\ref{al:ENKFFCNN}.

\begin{algorithm}
\caption{Algorithm for the EnKF-MLC framework}
\begin{algorithmic}[1]
\State {\bf{Input}}: $\boldsymbol{S}_{m,0}^{\mathfrak{N}}$ (initial conditions); $\mathcal{F}$ (pretrained function)
\State {\bf{Begin}}:\\
initialize:\\
\hspace{1.2cm}  $t=t_0,~j=0$, $\boldsymbol{S}_{a,0}^{\mathfrak{N}}=\boldsymbol{S}_{m,0}^{\mathfrak{N}}$\\
time loop:
\State \hspace{1.2cm} {\bf{while}} $t \leq t_{\text{max}} $ {\bf{do}}\\
\hspace{1.6cm} $\boldsymbol{S}^{\mathfrak{N}}_{f,j+1}=\mathcal{M}_{j+1:j}(\boldsymbol{S}^{\mathfrak{N}}_{a,j})$\\
\hspace{1.6cm} Calculate $\boldsymbol{P}^{\mathfrak{N}}_{f,j+1}$ with~\eqref{eq:covf}\\
\hspace{1.6cm} Predict covariance correction term $\Delta\boldsymbol{P}_{f,j+1}$ with Eq.~\eqref{eq:correction}\\
\hspace{1.6cm} Update $\boldsymbol{P}^{\mathfrak{N}}_{f,j+1}$ with Eq.~\eqref{eq:pcorrected}\\
\hspace{1.6cm} Resample forecast ensemble 
${\boldsymbol{S}}^{\mathfrak{N}}_{f,j+1}$ with  $\bar{\boldsymbol{S}}^{\mathfrak{N}}_{f,j+1}$ and ${\boldsymbol{P}}^{\mathfrak{N}}_{f,j+1}$\\
\hspace{1.6cm} Calculate $\boldsymbol{S}^{\mathfrak{N}}_{a,j+1}$ with Eq.~\eqref{eq:enkf}\\
\hspace{1.6cm}  $j \Leftarrow j+1$; $t \Leftarrow t_{j}$\\
\hspace{1.2cm} {\bf{end}} \\
{\bf{end}} 
\end{algorithmic}
\label{al:ENKFFCNN}
\end{algorithm}

\section{Numerical Experiments and Results}
\label{sec:Numerical_Experiments}

In this study, we evaluate the performance of the proposed EnKF-MLC framework through a series of numerical experiments based on the Lorenz systems, including both Lorenz-63 and Lorenz-96. In this section, we first present the results for two benchmark cases ($\S$~\ref{sec:Benchmark}), and then systematically demonstrate the feasibility and robustness of the proposed method under different configurations, by varying the ensemble size ($\S$~\ref{sec:Ensemble Size}), available observations ($\S$~\ref{sec:Observations}), and DA frequency ($\S$~\ref{sec:DA Frequency}).

\subsection{Benchmark Simulation Experiments}
\label{sec:Benchmark}

\subsubsection{Lorenz-63 system}
\label{sec:Lorenz-63}

The Lorenz-63 system is governed by the following set of ordinary differential equations
\begin{equation}
\begin{split}
     \frac{dx}{dt} &= \sigma \left( y - x\right),\\
     \frac{dy}{dt} &= x (\rho - z) - y,\\
     \frac{dz}{dt} &= xy - \beta z,
\end{split}
\label{eq:Lorenz-63 differential}
\end{equation}
where $x$, $y$, and $z$ represent the system state variables. The parameters $\sigma = 10$, $\rho = 28$, and $\beta = 8/3$ characterize the inherent properties of the dynamic system. The forward numerical integration is conducted with a fourth-order Runge–Kutta method and the numerical model time step is set to be $0.01$ model time units (i.e., $\Delta t = 0.01~\text{MTU}$).

To perform the DA operation, as shown in Eq.~\eqref{eq:enkf}, full or sparse noisy observations are needed. In this regard, we first run a reference simulation with presumed exact initial conditions, which results in the (synthetic) true state trajectory $\boldsymbol{s}_{T,j}$. Specifically, the exact initial condition is taken as the system state after a spin-up period of $200~\text{MTUs}$, in order to fully get rid of the transient effects. Then, a random error $\boldsymbol{\delta}$ drawn from a normal distribution with zero-mean and covariance $\boldsymbol{R}$ is added to ${\boldsymbol{s}}_{T,j}$, producing the synthetic noisy observation 
\begin{equation}
\bar{\boldsymbol{s}}_{m,j}={\boldsymbol{s}}_{T,j}+\boldsymbol{\delta}.
\end{equation}
Specifically, we set $\boldsymbol{R}=A \boldsymbol{I}$, where $A$ is the error magnitude (with $A=2$ in this study) and $\boldsymbol{I}$ denotes the identity matrix. Finally, the observation ensemble members are generated by adding random perturbations
\begin{equation}
    \boldsymbol{s}^{(n)}_{m,j}= \bar{\boldsymbol{s}}_{m,j}+\boldsymbol{\delta}^{(n)}_e,
\end{equation} 
where $\boldsymbol{\delta}^{(n)}_e$ is a random error following the same distribution as $\boldsymbol{\delta}$.

The datasets required to build the ML-based covariance correction function Eq.~\eqref{eq:correction} are generated by performing traditional EnKF simulations (without covariance correction) using both a large ensemble size $\mathcal{N}$ (which is set to be $100$ for the following numerical cases) and a small ensemble size $\mathfrak{N}$, for $100$ sets of initial conditions. The resulting data are partitioned into training, validation, and test sets with a ratio of $65:15:20$. During training, we utilize the rectified linear unit (ReLU) as the activation function and the mean squared error (MSE) as the loss function. 

\begin{figure}
    \centering
    \begin{subfigure}[b]{0.32\linewidth}
        \centering
        \includegraphics[width=\linewidth]{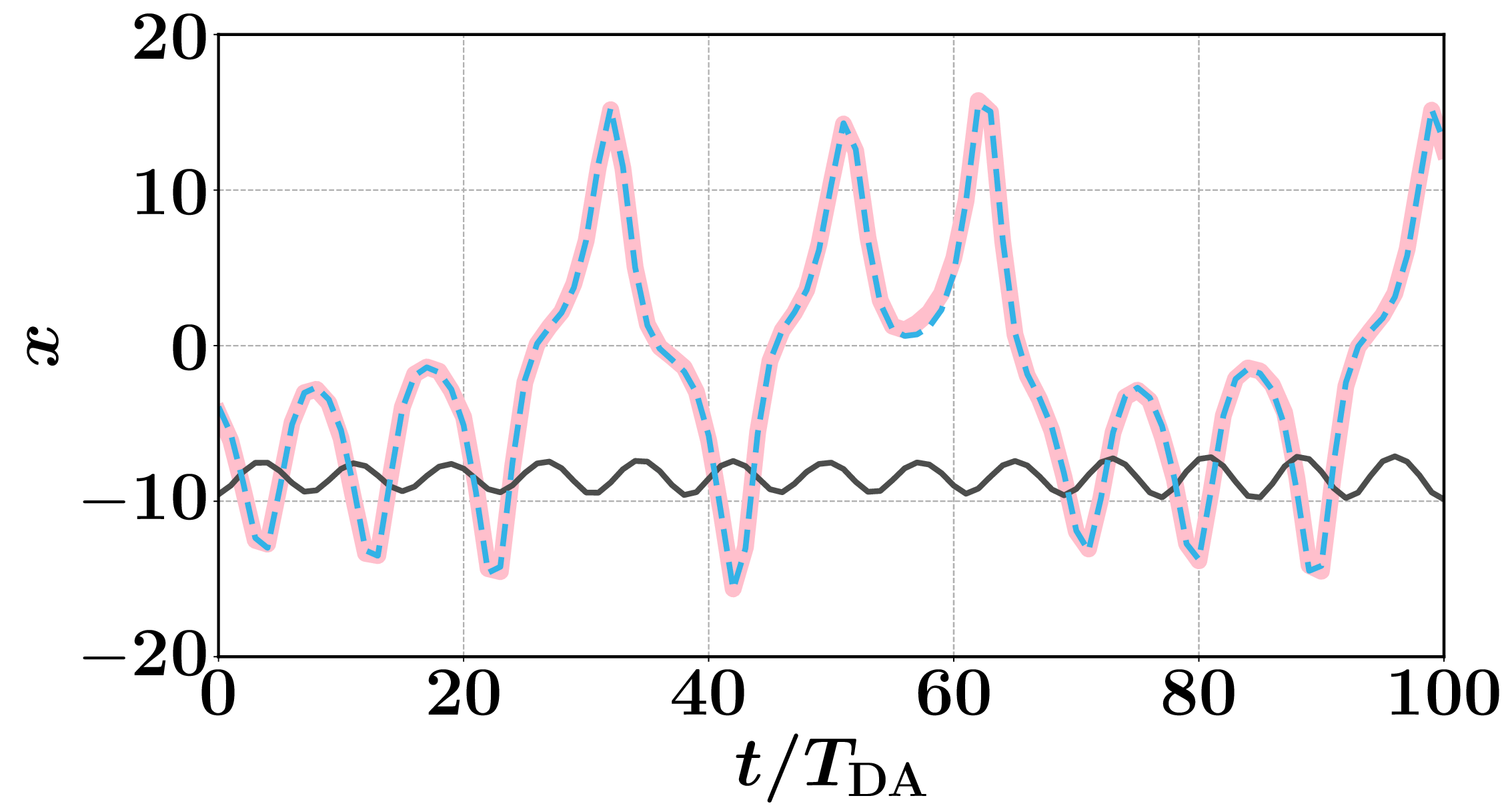}
        \caption{}
        \label{fig:sub1}
    \end{subfigure}
    \hfill
    \begin{subfigure}[b]{0.32\linewidth}
        \centering
        \includegraphics[width=\linewidth]{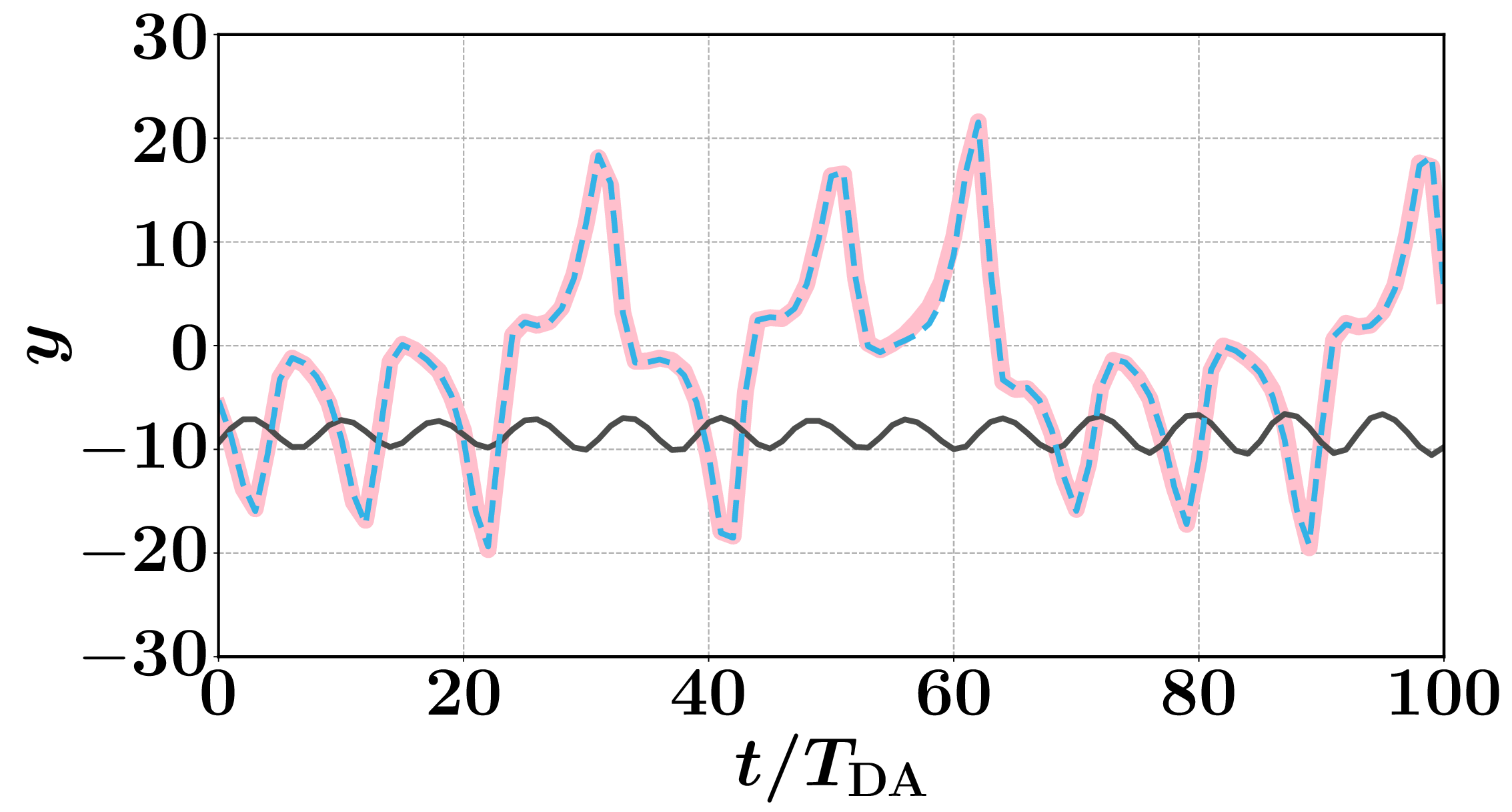}
        \caption{}
        \label{fig:sub2}
    \end{subfigure}
    \hfill
    \begin{subfigure}[b]{0.32\linewidth}
        \centering
        \includegraphics[width=\linewidth]{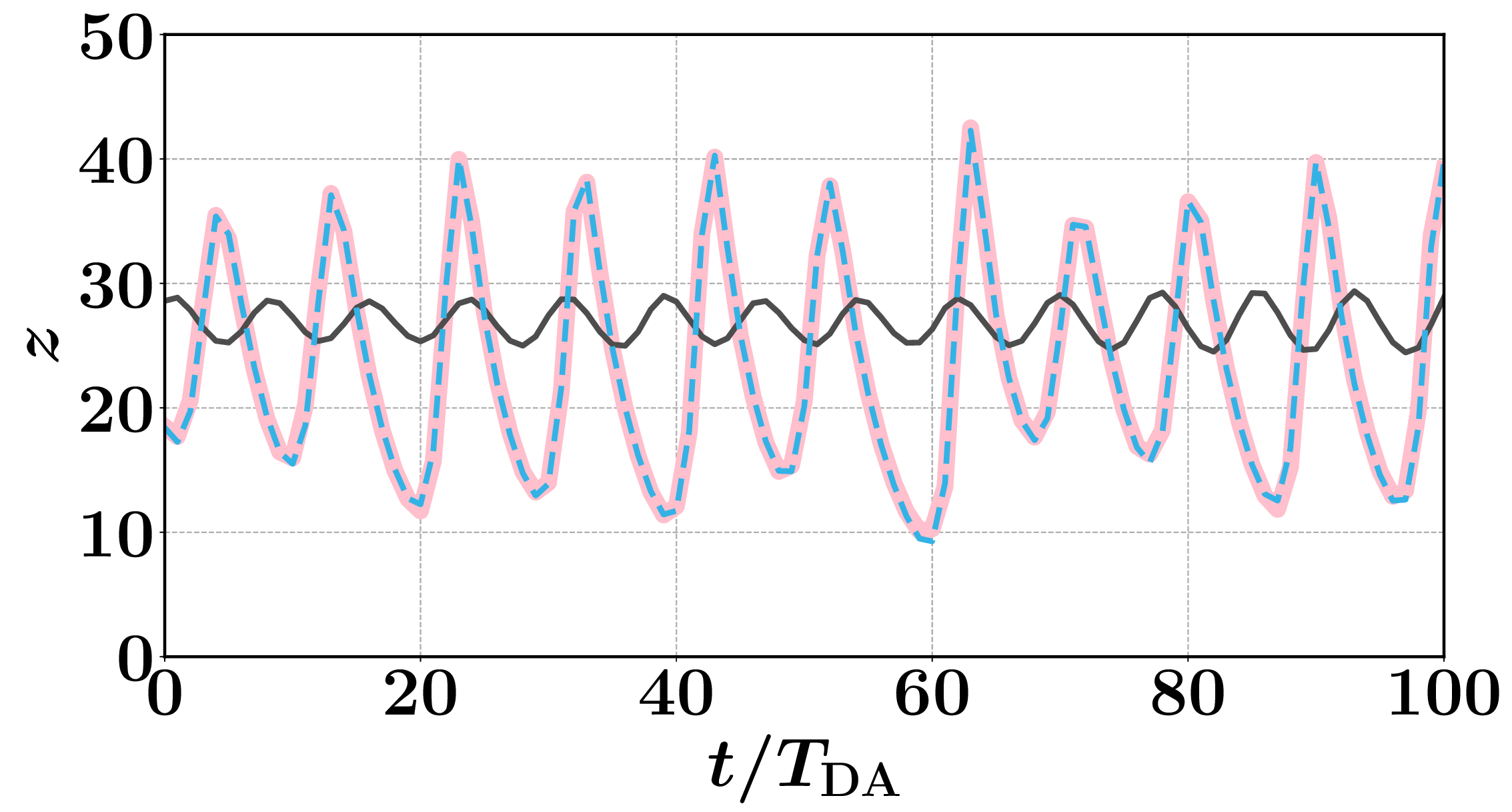}
        \caption{}
        \label{fig:sub3}
    \end{subfigure}
    
    \caption{Analysis results obtained with the traditional EnKF using $\mathcal{N}=100$ ({\color{mypink}\rule[0.5ex]{0.5cm}{2.0pt}}) and $\mathfrak{N}=3$ ({\color{darkgray}\rule[0.5ex]{0.5cm}{0.5pt}}), as well as the true solution ({\color{cyan}\dashL}), for Lorenz-63 benchmark case: (a)~$x$, (b)~$y$, and (c)~$z$.}
    \label{fig:L63state_xyz8dt}
\end{figure}

The benchmark experiment is conducted with $\mathfrak{N}=3$, a DA interval $T_{\text{DA}}=0.08~\text{MTU}$, and all three state variables observed. Fig.~\ref{fig:L63state_xyz8dt} shows the analysis results from the traditional EnKF using both $\mathcal{N}=100$ and $\mathfrak{N}=3$, as well as the truth. It can be clearly observed that, with $\mathcal{N}=100$, the analysis can accurately capture the true state trajectory. However, when the ensemble size is reduced to $\mathfrak{N}=3$, a significant discrepancy between the analysis and truth shows up, which is mainly due to the inaccurate estimation of the forecast covariances.

\begin{figure}
    \centering
    \begin{subfigure}[b]{0.32\linewidth}
        \centering
        \includegraphics[width=\linewidth]{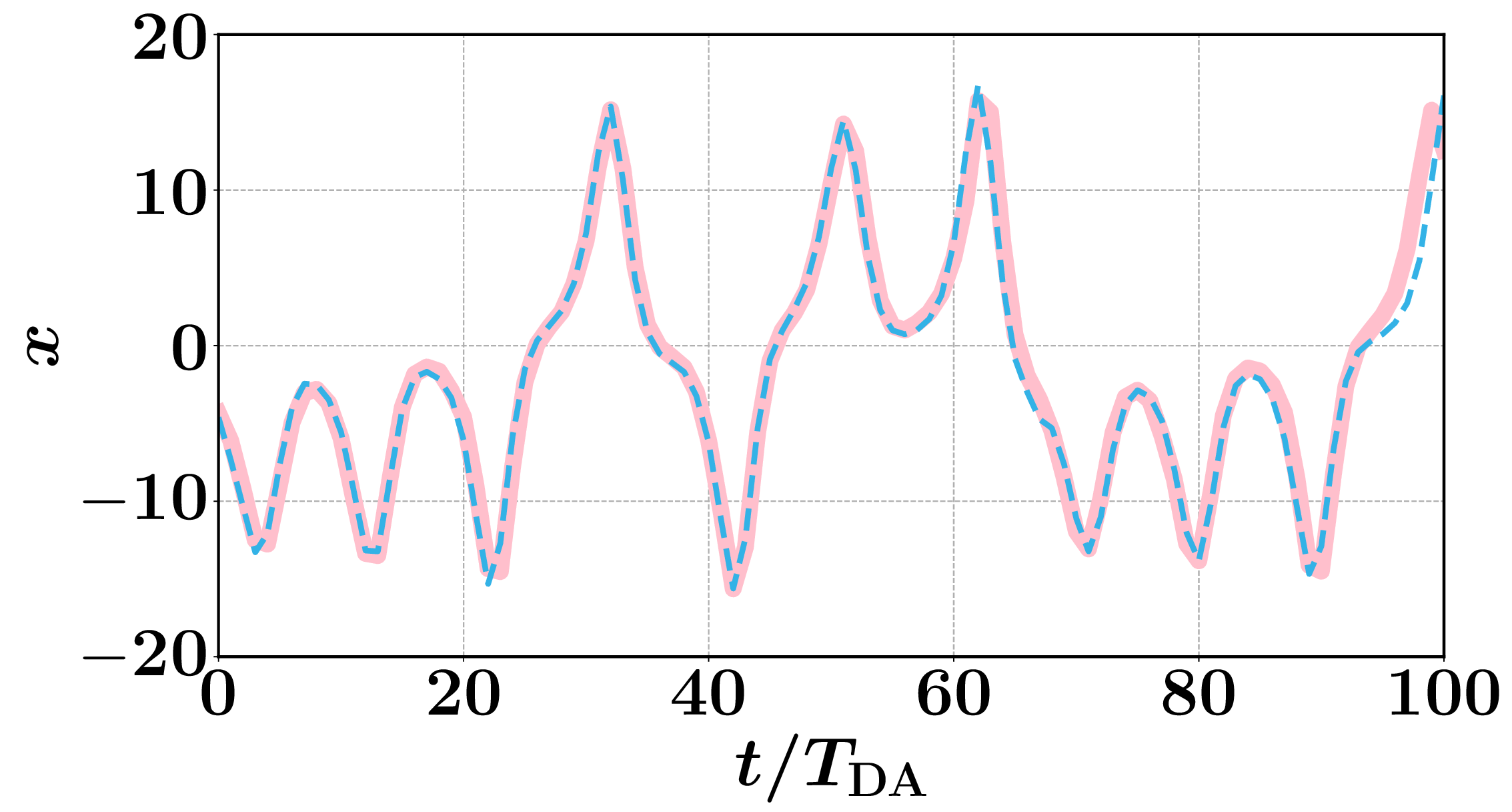}
        \caption{}
        \label{fig:sub1}
    \end{subfigure}
    \hfill
    \begin{subfigure}[b]{0.32\linewidth}
        \centering
        \includegraphics[width=\linewidth]{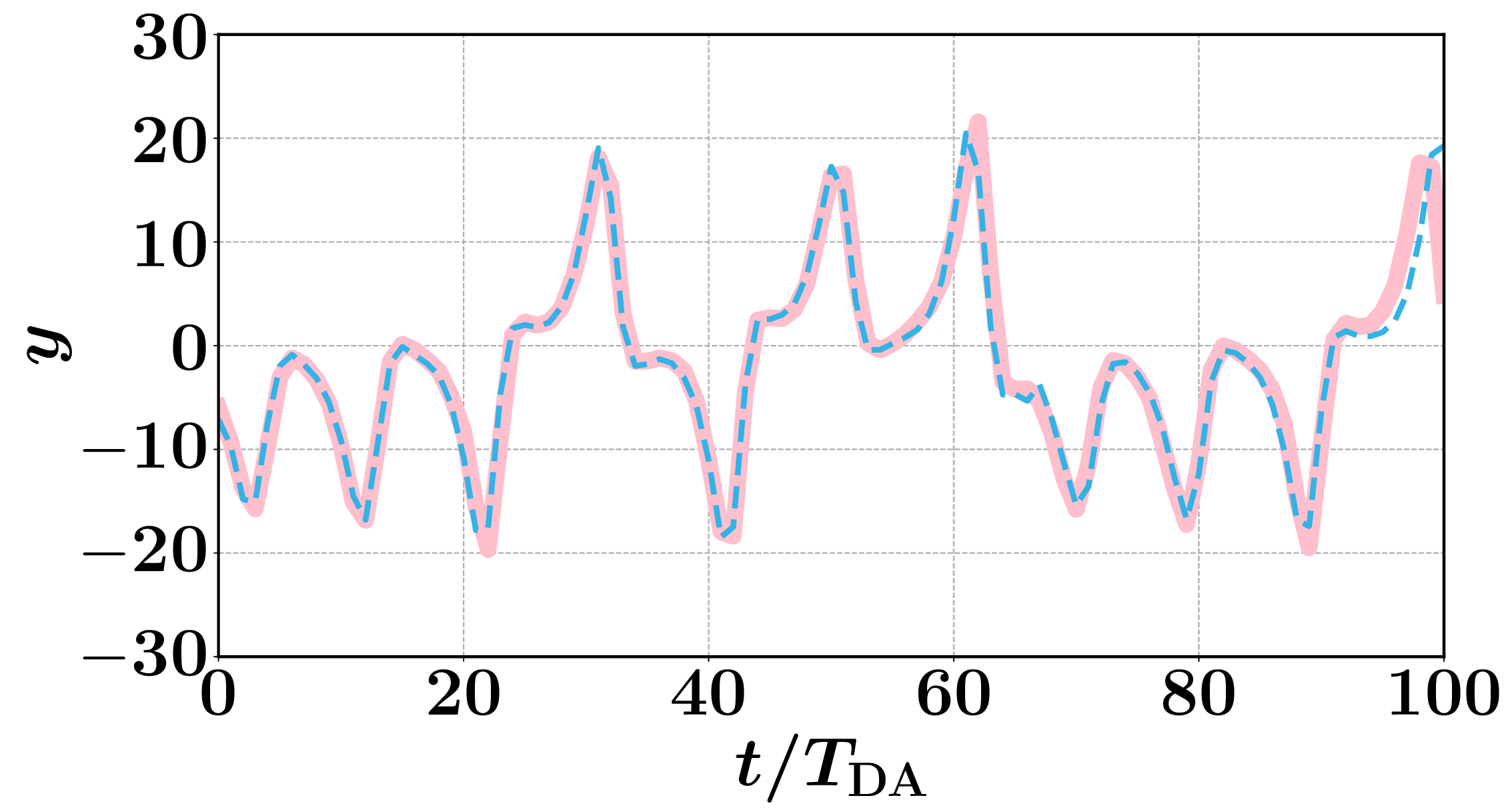}
        \caption{}
        \label{fig:sub2}
    \end{subfigure}
    \hfill
    \begin{subfigure}[b]{0.32\linewidth}
        \centering
        \includegraphics[width=\linewidth]{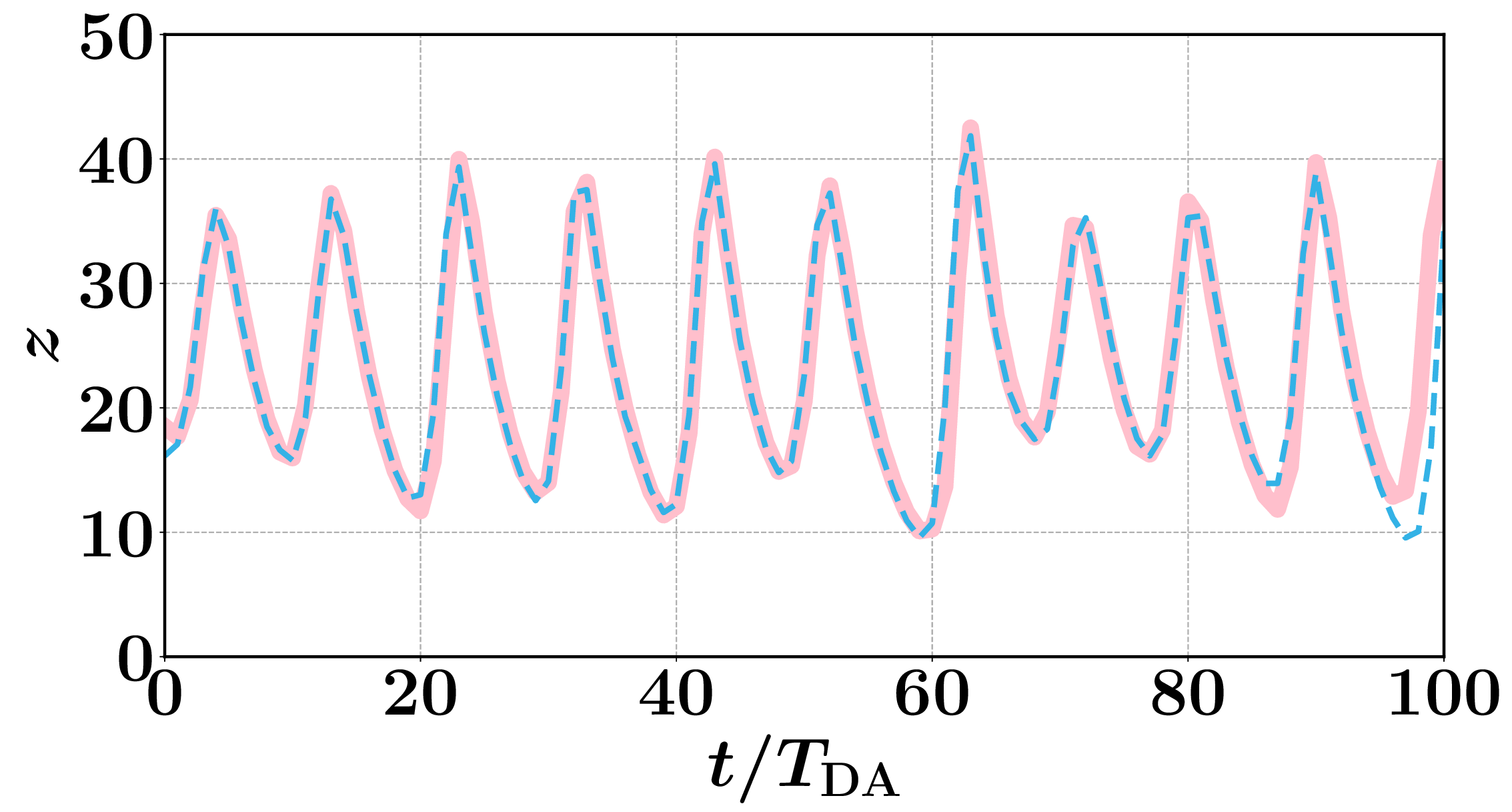}
        \caption{}
        \label{fig:sub3}
    \end{subfigure}
    
    \caption{Analysis results from the proposed EnKF-MLC framework with $\mathfrak{N}=3$ ({\color{cyan}\dashL}) and the traditional EnKF with $\mathcal{N}=100$ ({\color{mypink}\rule[0.5ex]{0.5cm}{2.0pt}}): (a)~$x$, (b)~$y$, and (c)~$z$.}
    \label{fig:L63stateonline}
\end{figure}

The proposed EnKF-MLC framework is then applied to this benchmark case. Fig.~\ref{fig:L63stateonline} presents the analysis results produced by the EnKF-MLC method with $\mathfrak{N}=3$ and traditional EnKF with $\mathcal{N}=100$, with the former closely following the latter. This remarkable improvement confirms that the proposed EnKF-MLC algorithm can effectively address the degraded uncertainty quantification induced by a limited ensemble size, offering a computationally efficient alternative without sacrificing accuracy.

\begin{figure}
    \centering
    \begin{subfigure}[h]{0.49\textwidth}
        \centering   \includegraphics[width=\textwidth]{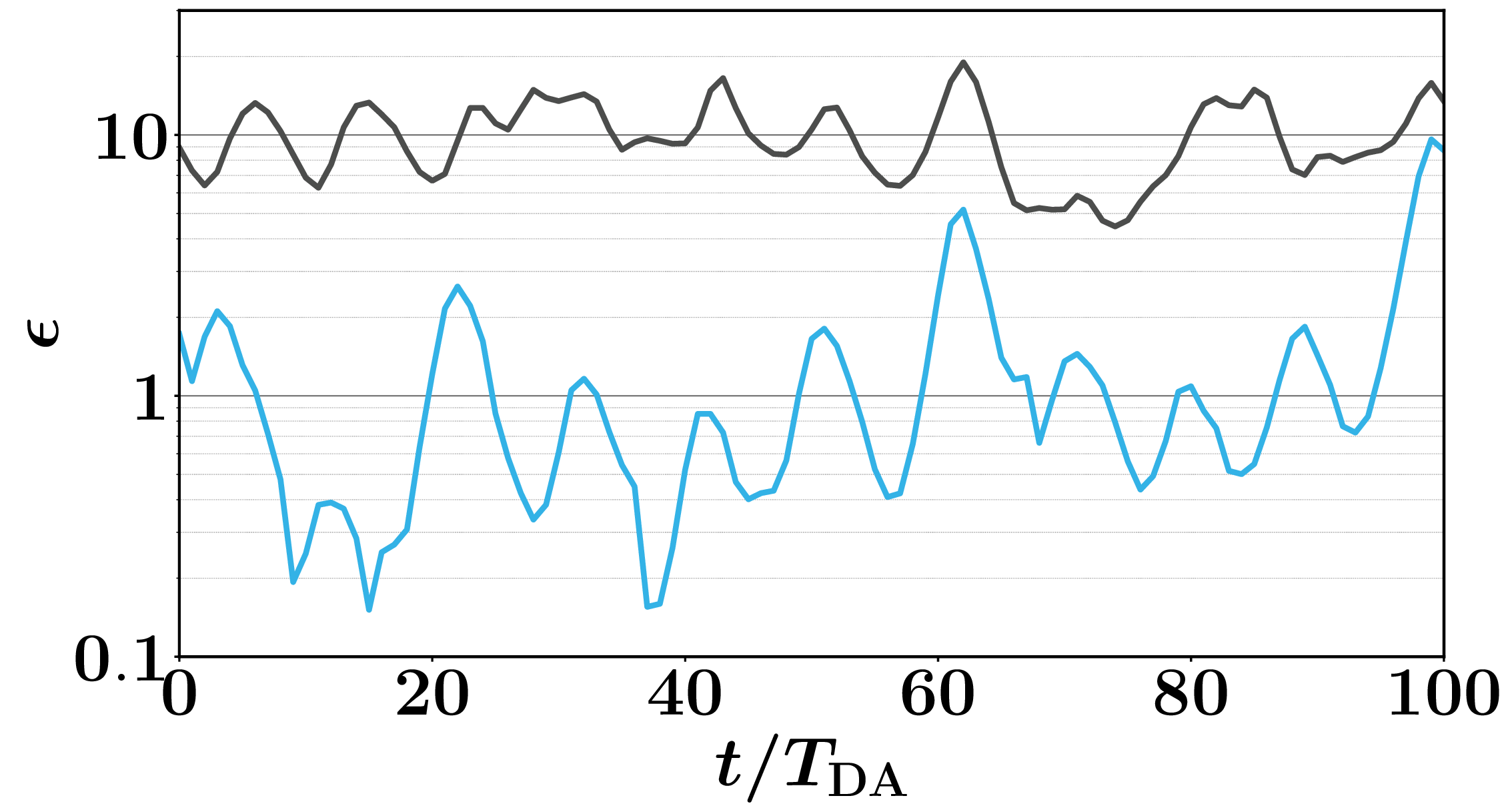}
        \caption{}  
        \label{fig:L63error}
    \end{subfigure}  
    \hfill
    \begin{subfigure}[h]{0.49\textwidth}
        \centering     \includegraphics[width=\textwidth]{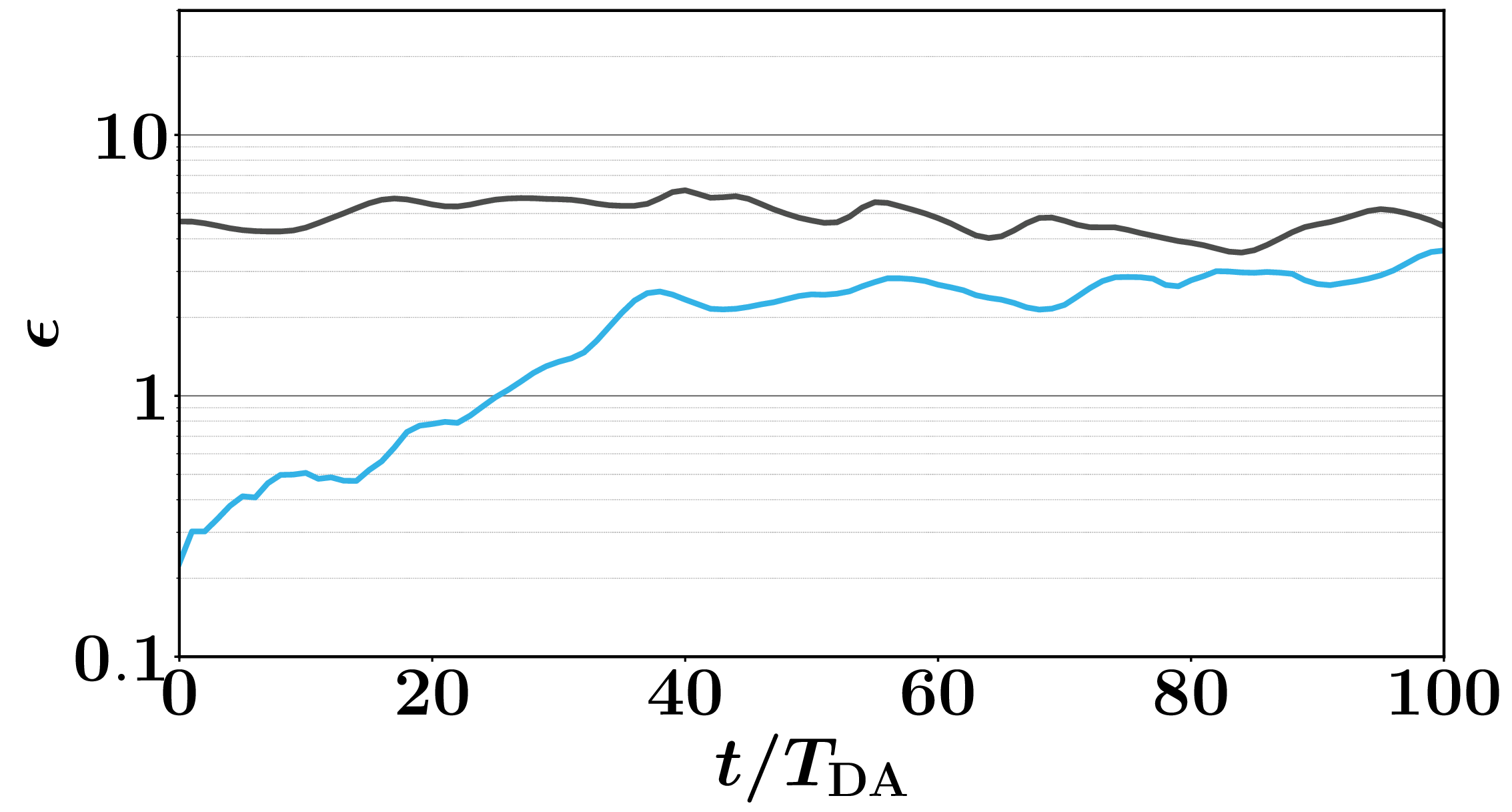}      
        \caption{}  
        \label{fig:L96error}
    \end{subfigure}
   \caption{Time histories of $\epsilon$ with traditional EnKF ({\color{darkgray}\rule[0.5ex]{0.5cm}{0.5pt}}) and EnKF-MLC ({\color{cyan}\rule[0.5ex]{0.5cm}{0.5pt}}) for benchmark cases:  (a) Lorenz-63 and (b) Lorenz-96.}
    \label{fig:epsilonhis}
 \end{figure}

In addition, an error metric is defined to quantitatively evaluate the overall performance of the proposed EnKF-MLC framework
\begin{equation}    
\epsilon(t_j)=\sqrt{\frac{1}{K}\Sigma_{k=1}^{K}\left(\bar{\boldsymbol{s}}^{\mathfrak{N}}_{a,j,k}-\bar{\boldsymbol{s}}^{\mathcal{N}}_{a,j,k}\right)^{2}},
\label{eq:L2norm}
\end{equation}
where $K$ is the number of initial conditions in the test dataset, with $k$ being the index.  Here we evaluate $\epsilon(t_j)$ using both the traditional EnKF and EnKF-MLC with the same limited ensemble size $\mathfrak{N}=3$, for which the results are shown in Fig.~\ref{fig:epsilonhis}(a). It can be found that, after applying the proposed EnKF-MLC framework, $\epsilon$ is reduced by about one order of magnitude, although it exhibits a slight upward trend due to the accumulation of residual errors in Eq.~\eqref{eq:correction}.

\subsubsection{Lorenz-96 system}
\label{sec:Lorenz-96}
To further evaluate the scalability and robustness of the proposed EnKF-MLC framework, we extend the numerical experiments to the Lorenz-96 system, which is formulated as
\begin{equation}
\label{eq:Lorenz-96 differential}
\frac{dx^{(i)}}{dt} = (x^{(i+1)} - x^{(i-2)}) x^{(i-1)} - x^{(i)} + F, i=1,2\dots L,
\end{equation}
where $L$ is the total number of variables and $x^{(i)}$ denotes the $i$-th state variable. In this study, we consider $L=40$ and impose periodic boundary conditions. The constant external forcing term is set as $F=8$. The time integration of Eq.~\eqref{eq:Lorenz-96 differential} is performed in the same manner as Eq.~\eqref{eq:Lorenz-63 differential}, i.e., with the fourth-order Runge–Kutta method and $\Delta t=0.01~\text{MTU}$.

\begin{figure}
    \centering
    \begin{subfigure}[b]{0.32\linewidth}
        \centering
        \includegraphics[width=\linewidth]{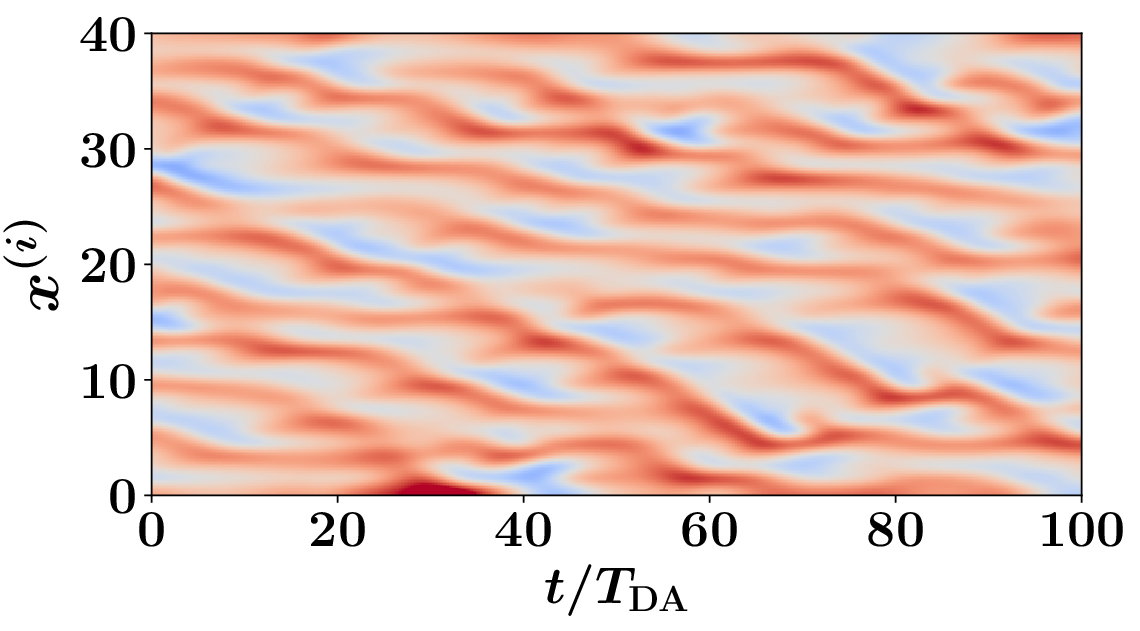}
        \caption{}
        \label{fig:sub1}
    \end{subfigure}
    \hfill
    \begin{subfigure}[b]{0.32\linewidth}
        \centering
        \includegraphics[width=\linewidth]{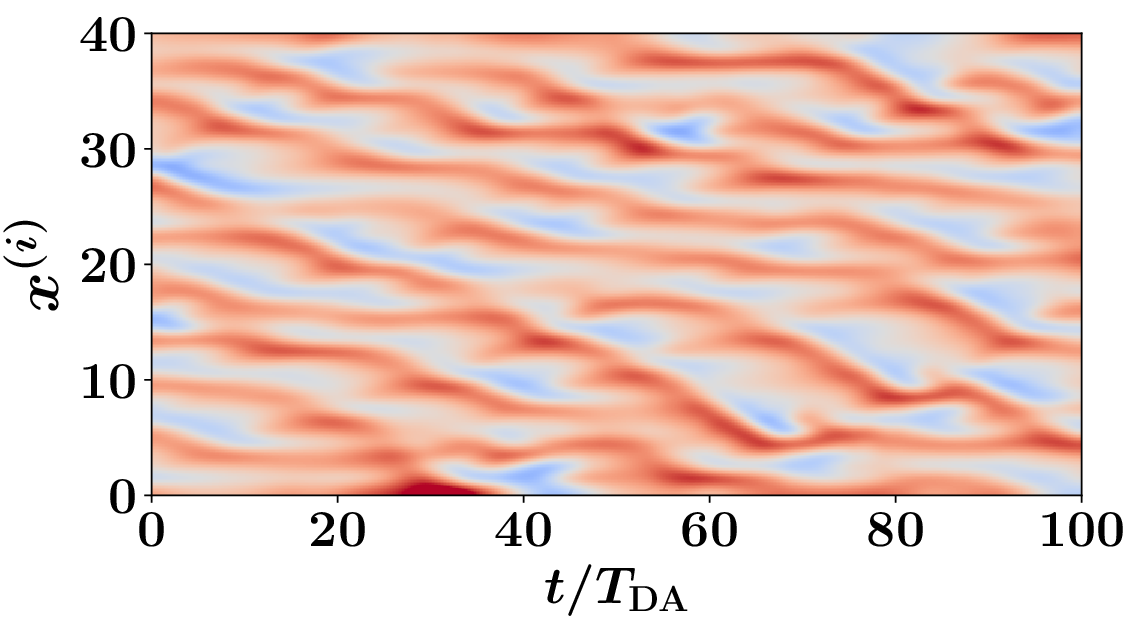}
        \caption{}
        \label{fig:sub2}
    \end{subfigure}
    \hfill
    \begin{subfigure}[b]{0.32\linewidth}
        \centering
        \includegraphics[width=\linewidth]{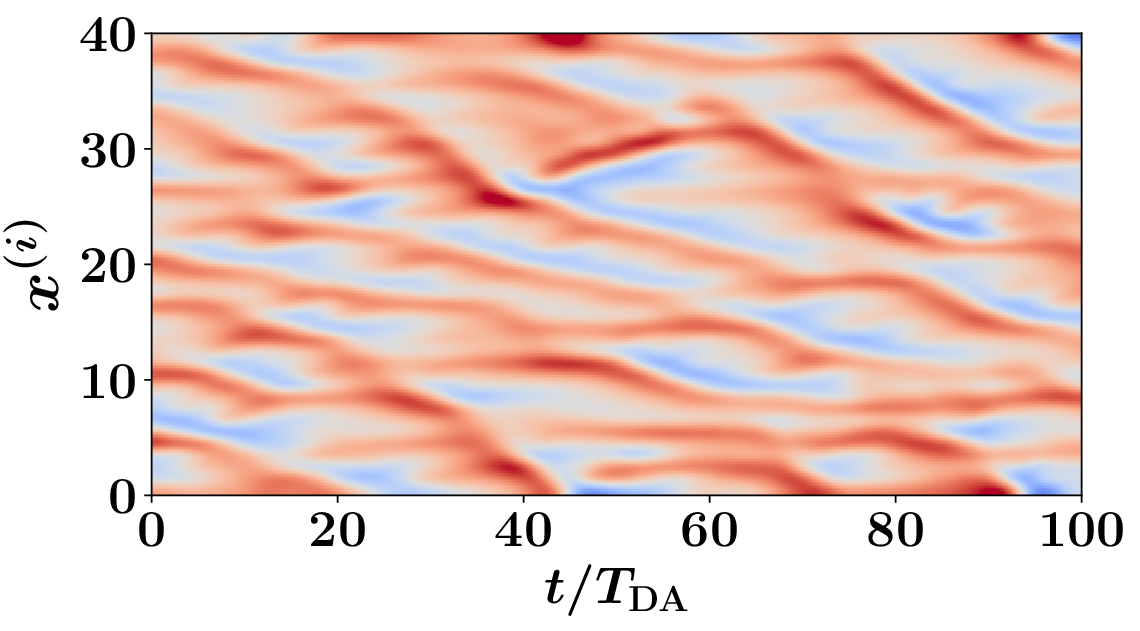}
        \caption{}
        \label{fig:sub3}
    \end{subfigure}
    
    \vspace{0.05cm}  
    \includegraphics[width=0.4\linewidth]{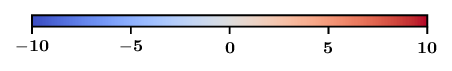}  
    
    \caption{Analysis results obtained with the true solution (a), the traditional EnKF using $\mathcal{N}=100$ (b) and $\mathfrak{N}=10$ (c) for Lorenz-96 benchmark case.}
    \label{fig:L96state_40obs5dtN10}
\end{figure}

For the benchmark configuration, we adopt an ensemble size $\mathfrak{N}=10$, an assimilation interval $T_{\text{DA}}=0.05~\text{MTU}$, and observations of all $40$ variables, i.e., $\{x^{(i)} \mid i=1,2,\dots,40\}$. A large ensemble with $\mathcal{N}=100$ is first used to define the correction term in Eq.~\eqref{eq:correct}. Moreover, both inflation and localization are applied to ensure numerical stability and enable the analysis to closely follow the true state trajectory (see Fig.~\ref{fig:L96state_40obs5dtN10}). Specifically, a constant inflation factor $\lambda=1.01$ and the Gaspari-Cohn localization function~\citep{gaspari1999construction,carrassi2018data} with the localization radius $\mathcal{R}=40$ are adopted.

\begin{figure}
    \centering
    \begin{subfigure}[b]{0.32\linewidth}
        \centering
        \includegraphics[width=\linewidth]{L96state_100_Hovmoller.eps}
        \caption{}
        \label{fig:sub1}
    \end{subfigure}
    \hfill
    \begin{subfigure}[b]{0.32\linewidth}
        \centering
        \includegraphics[width=\linewidth]{L96state_10_Hovmoller.eps}
        \caption{}
        \label{fig:sub2}
    \end{subfigure}
    \hfill
    \begin{subfigure}[b]{0.32\linewidth}
        \centering
        \includegraphics[width=\linewidth]{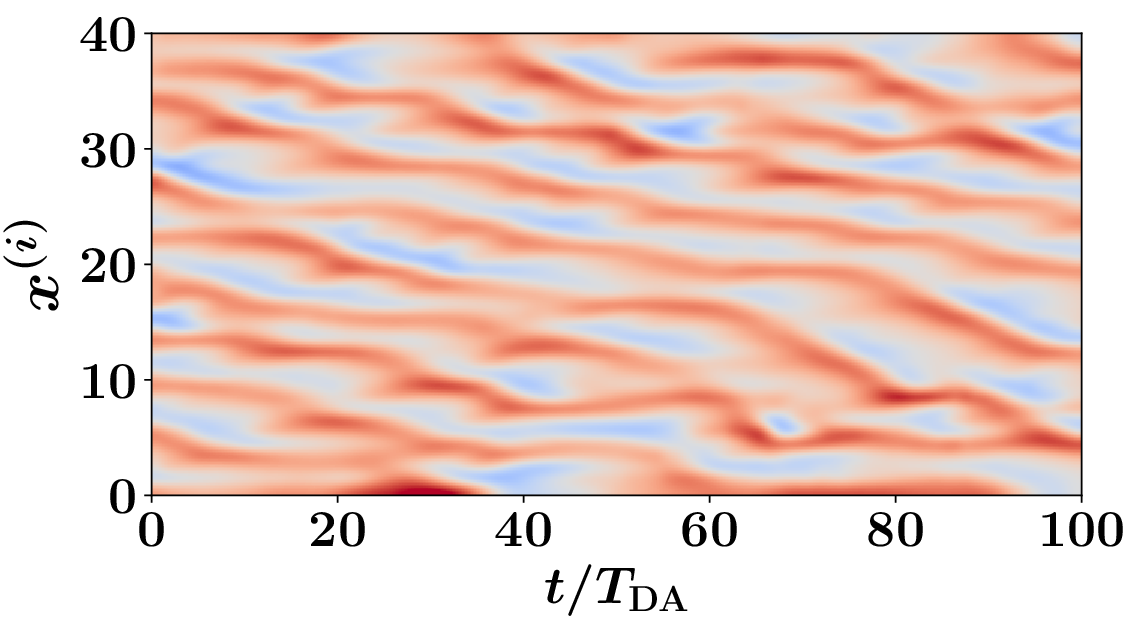}
        \caption{}
        \label{fig:sub3}
    \end{subfigure}
    
    \vspace{0.05cm}  
    \includegraphics[width=0.4\linewidth]{colorbar10.eps}  
    
    \caption{Analysis results given by the traditional EnKF using $\mathcal{N}=100$ (a) and $\mathfrak{N}=10$ (b), as well as the EnKF-MLC algorithm with $\mathfrak{N}=10$ (c) for Lorenz-96 benchmark case.}
    \label{fig:L96time_40obs5dtN10}
\end{figure}

For the small ensemble case, we first examine the results given by the traditional EnKF algorithm. As illustrated in Fig.~\ref{fig:L96state_40obs5dtN10}, the traditional EnKF fails to track the true state trajectory with such a small ensemble size. However, as shown in Fig.~\ref{fig:L96time_40obs5dtN10}, when the MLP function is applied to amend the forecast statistics at each analysis step, a remarkable improvement is observed, which confirms the effectiveness of the proposed method in mitigating the issues caused by a limited ensemble size for high-dimensional systems.

For Lorenz‑96, $\epsilon(t_j)$ defined in Eq.~\eqref{eq:L2norm} is again used to further assess the performance of the proposed EnKF‑MLC framework. The time evolution of $\epsilon$ for both the traditional EnKF and the EnKF‑MLC is presented in Fig.~\ref{fig:epsilonhis}(b). It is observed that the EnKF‑MLC consistently achieves a substantially lower value of $\epsilon$ throughout the entire simulation period compared to the traditional EnKF. In particular, $\epsilon$ is reduced more than one order of magnitude during the early stage. However, it should be noted that as the simulation propagates, $\epsilon$ given by the EnKF‑MLC algorithm shows an expansion trend, which is similar to yet more pronounced than that observed for Lorenz-63. This greater error growth rate in Lorenz-96 is mainly due to its higher dimensionality and stronger nonlinearity, which amplify the residual error accumulation.

Finally, we evaluate the computational time of running the EnKF-MLC algorithm as compared to that of performing a single simulation for $T_\text{DA}$. As shown in Tab.~\ref{tab:CPU-time}, for Lorenz-63 the EnKF-MLC algorithm requires approximately one order of magnitude less computational time than a single simulation over one DA interval. For Lorenz-96, this computational efficiency advantage grows to two orders of magnitude, which means that the additional computational cost induced by the covariance correction operation is practically negligible.
 \begin{table}[h]
\centering
\caption{Computational time for different calculations on a single Intel Core Ultra 9 CPU}
\label{tab:CPU-time}
\begin{tabular}{ccc}
\toprule \textbf{Cases}
 & \textbf{A single simulation for $\boldsymbol{T}_\text{DA}$ (s)} & \textbf{EnKF-MLC (s)}\\ \midrule
Lorenz-63 & 3.29e-3 & 3.28e-4
 \\
Lorenz-96 & 3.93e-2
 & 3.86e-4
 \\
\bottomrule
\end{tabular}
\end{table}

\subsection{Sensitivity to Ensemble Size}
\label{sec:Ensemble Size}
\begin{figure}
    \centering
    \begin{subfigure}[b]{0.49\linewidth}
        \centering
        \includegraphics[width=\linewidth]{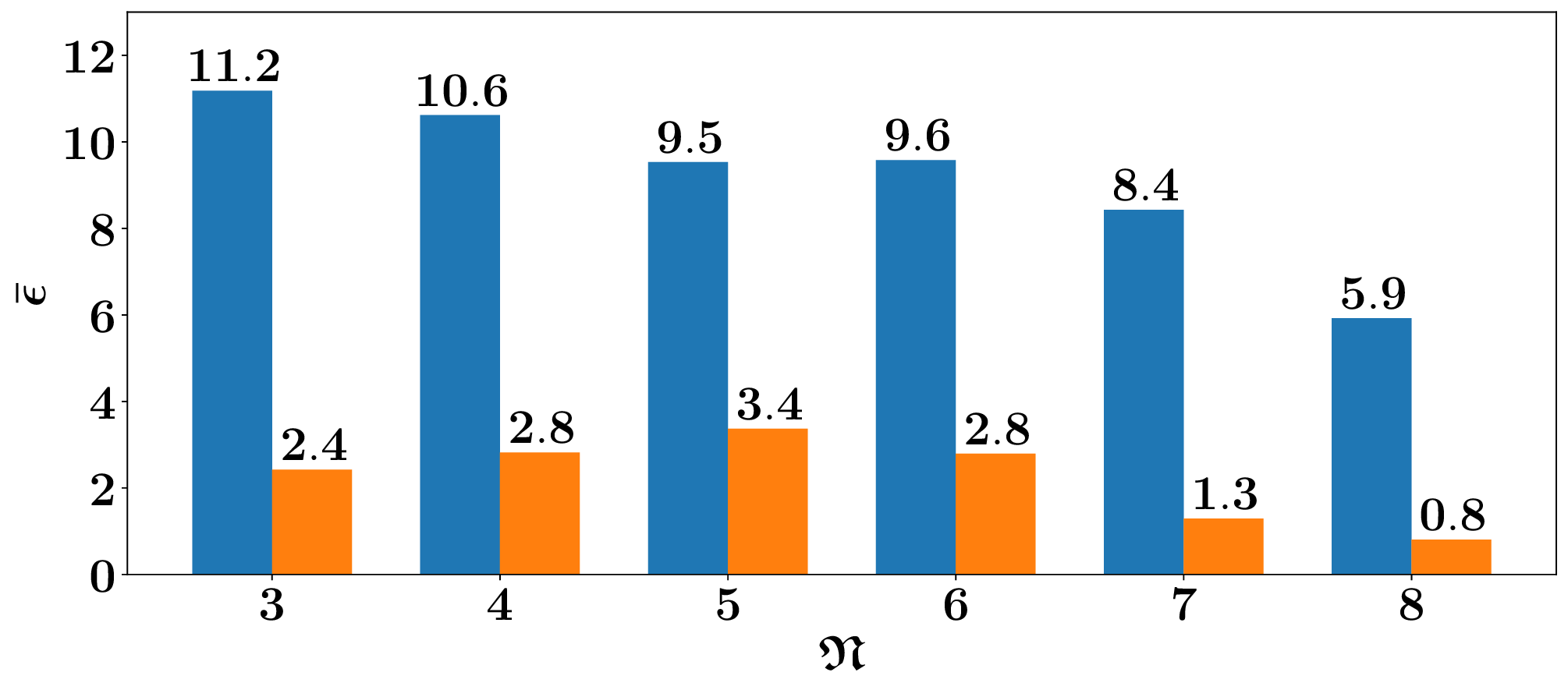}
        \caption{}
        \label{fig:L63_rmse_ensemble}
    \end{subfigure}
    \hfill
    \begin{subfigure}[b]{0.325\linewidth}
        \centering
        \includegraphics[width=\linewidth]{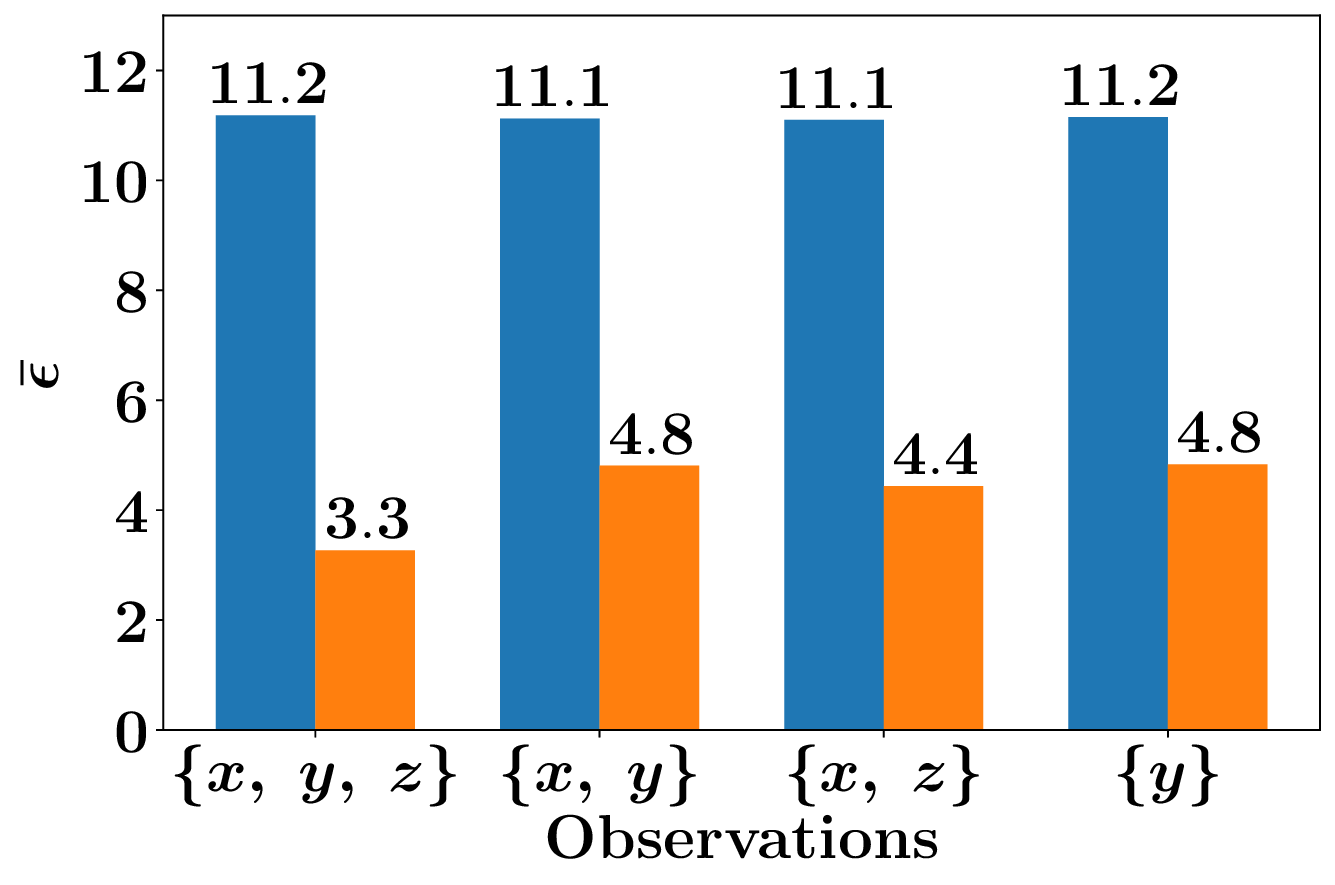}
        \caption{}
        \label{fig:L63_rmse_obs}
    \end{subfigure}
    \hfill
    \begin{subfigure}[b]{0.161\linewidth}
        \centering
        \includegraphics[width=\linewidth]{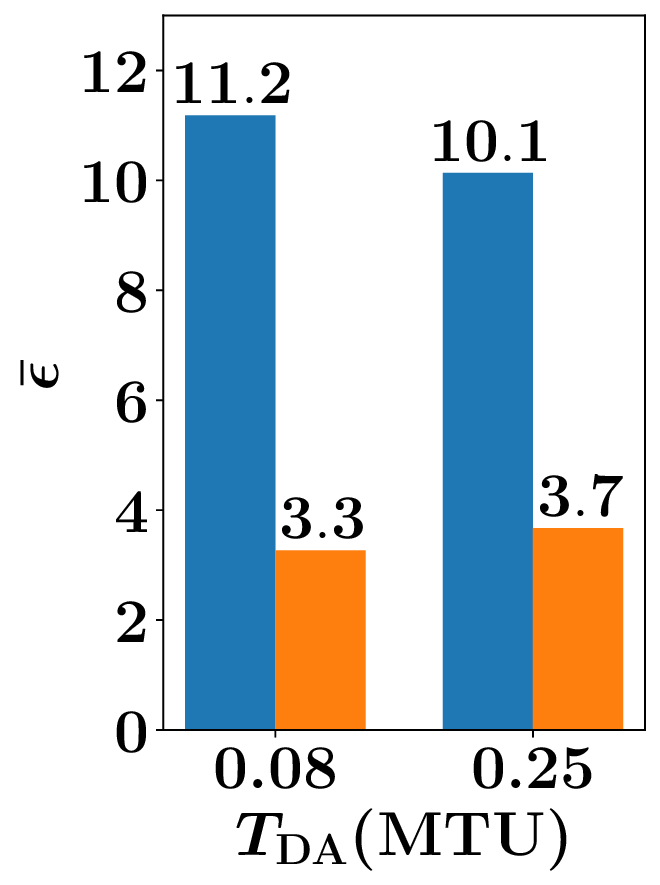}
        \caption{}
        \label{fig:L63_rmse_dt}
    \end{subfigure}
    \caption{$\bar\epsilon$ of the analysis for the Lorenz-63 system using the traditional EnKF with small ensemble size $\mathfrak{N}$ (\colorSquare{EnkfColor}) and the proposed EnKF-MLC framework (\colorSquare{EnKFMLCcolor}), evaluated across different (a)~ensemble sizes ($\mathfrak{N}=3,4,\dots,8$), (b)~available observations ($\{x,~y,~z\}$, $\{x,~y\}$, $\{x,~z\}$, and $\{y\}$
    ), and (c)~DA frequency ($T_{\text{DA}}=0.08~\text{and}~0.25~\text{MTU}$).}
    \label{fig:L63_rmse}
\end{figure}

To further assess the feasibility and robustness of the proposed EnKF-MLC framework, we first systematically evaluate its performance across different ensemble sizes $\mathfrak{N}$. For Lorenz-63, $\mathfrak{N}$ is varied between $3$ and $8$, while maintaining all other experimental configurations identical to the benchmark case. Here we consider one time-averaged error metric
\begin{equation}
    \bar{\epsilon}=\frac{1}{\mathcal{K}}\Sigma_{j=1}^{\mathcal{K}}\epsilon(t_j),
\end{equation}
where $\mathcal{K}$ is the number of time instances. The results are presented in Fig.~\ref{fig:L63_rmse}(a) for both the traditional EnKF and proposed EnKF-MLC methods. It is evident that for all ensemble sizes considered, the EnKF-MLC framework can consistently outperform the traditional EnKF by achieving significantly lower error levels. Specifically, the most significant improvement in terms of the relative error drop is observed for $\mathfrak{N}=8$, with $\bar\epsilon$ reduced by $86\%$.

\begin{figure}
    \centering
    \begin{subfigure}[b]{0.48\linewidth}
        \centering
        \includegraphics[width=\linewidth]{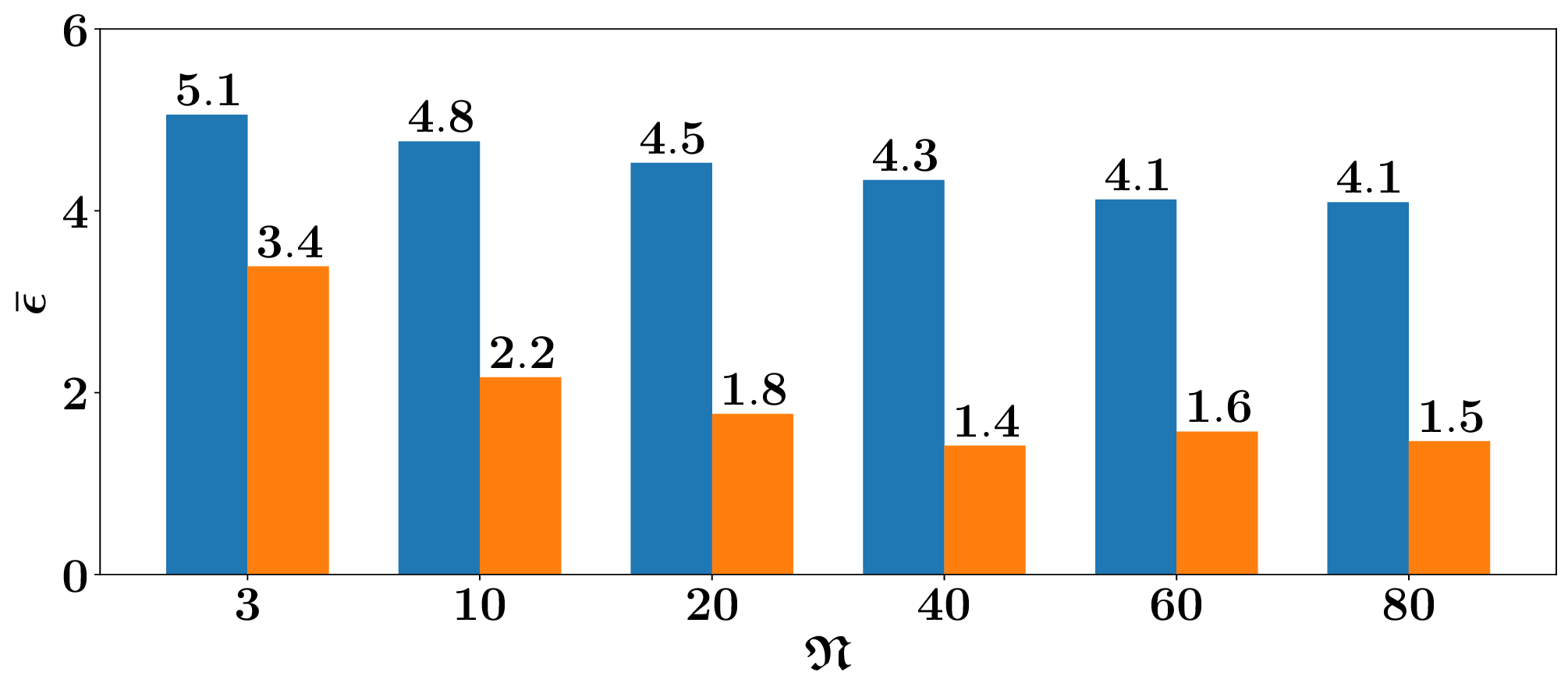}
        \caption{}
        \label{fig:L96_rmse_ensemble}
    \end{subfigure}
    \hfill
    \begin{subfigure}[b]{0.245\linewidth}
        \centering
        \includegraphics[width=\linewidth]{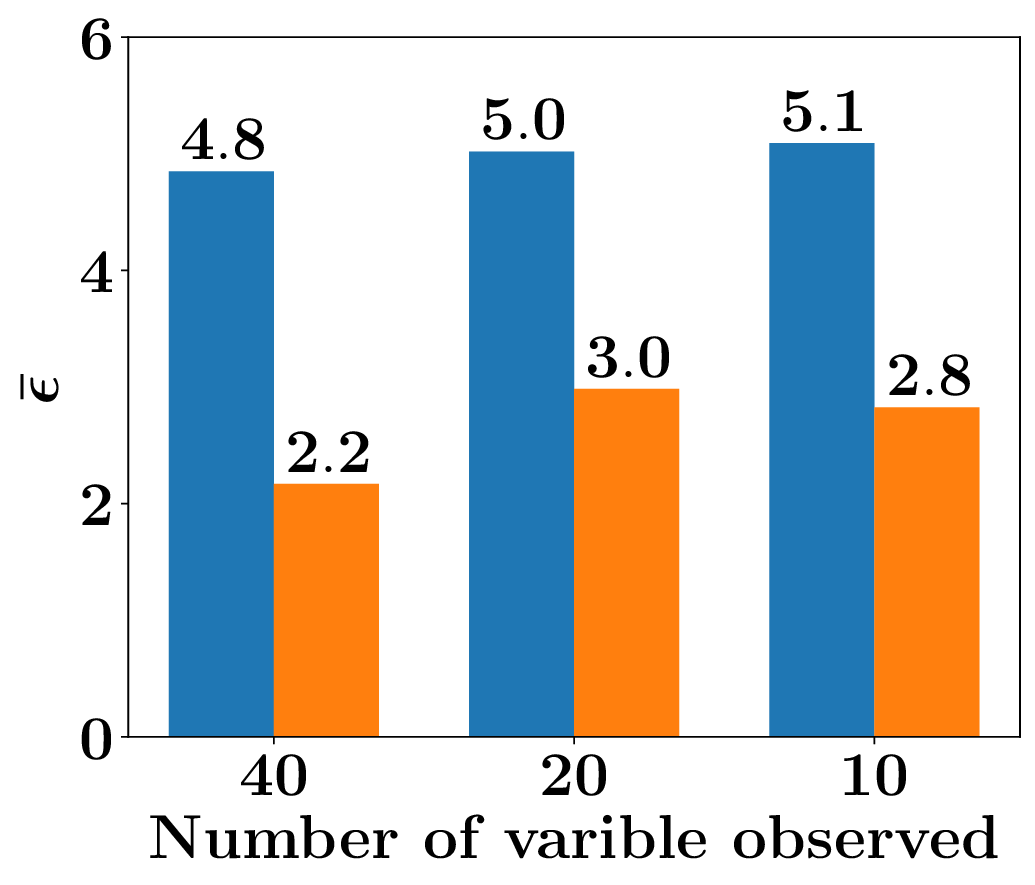}
        \caption{}
        \label{fig:L96_rmse_obs}
    \end{subfigure}
    \hfill
    \begin{subfigure}[b]{0.245\linewidth}
        \centering
        \includegraphics[width=\linewidth]{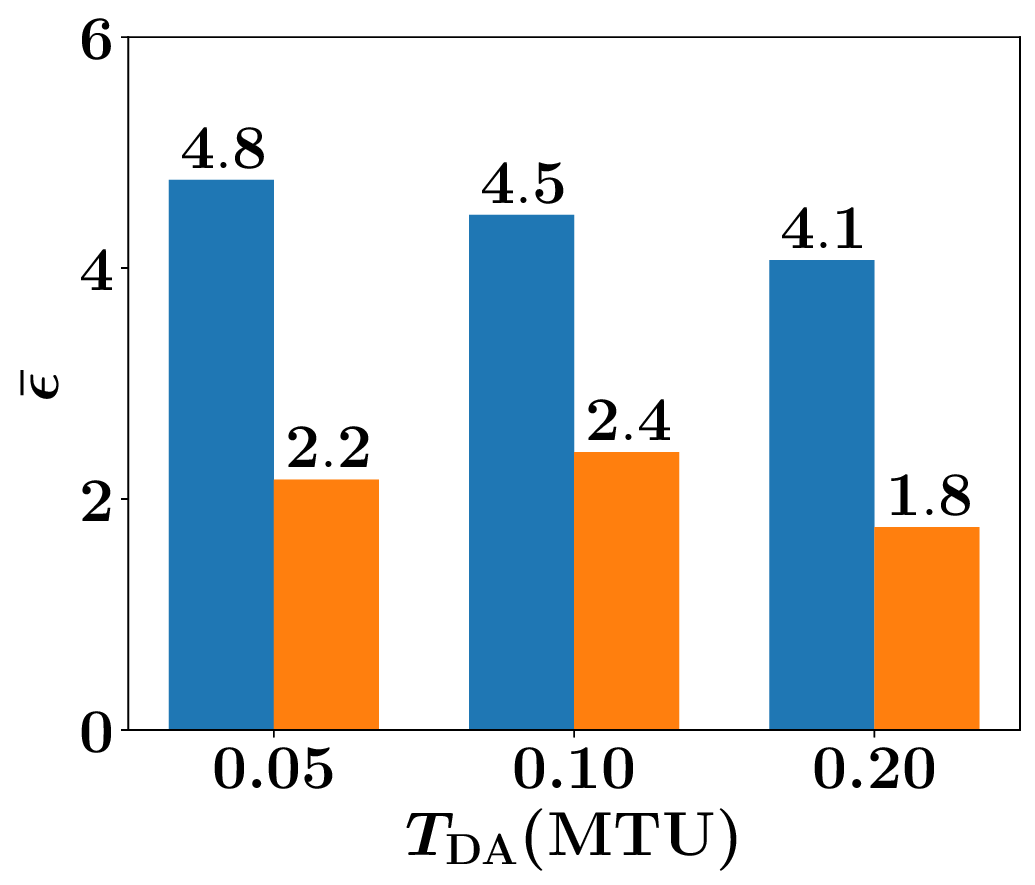}
        \caption{}
        \label{fig:L96_rmse_dt}
    \end{subfigure}
    \caption{$\bar\epsilon$ of the analysis for the Lorenz-96 system using the traditional EnKF with small ensemble size $\mathfrak{N}$ (\colorSquare{EnkfColor}) and the proposed EnKF-MLC framework (\colorSquare{EnKFMLCcolor}), evaluated across different (a)~ensemble sizes, (b)~available observations, and (c)~DA frequency.}
    \label{fig:L96_rmse}
\end{figure}

For Lorenz-96, we also investigate the impact of ensemble size $\mathfrak{N}$ by varying it from $3$ to $80$. The results of $\bar\epsilon$ for the Lorenz-96 system are presented in Fig.~\ref{fig:L96_rmse}(a). As the ensemble size $\mathfrak{N}$ increases from $3$ to $80$, the traditional EnKF shows a gradual decrease trend, but even at $\mathfrak{N}=80$ $\bar\epsilon$ still remains relatively high ($\bar\epsilon=4.1$). In contrast, the EnKF-MLC framework achieves dramatically lower values of $\bar\epsilon$ across all ensemble sizes. For the most challenging case $\mathfrak{N}=3$, the traditional EnKF yields $\bar\epsilon=5.1$, while EnKF-MLC reduces it to $3.4$, corresponding to a reduction of approximately $35\%$. The best improvement is achieved with $\mathfrak{N}=40$, where $\bar\epsilon$ drops from $4.3$ (traditional EnKF) to $1.4$ (EnKF-MLC) corresponding to a reduction of approximately $70\%$. This is likely because $\mathfrak{N}=40$ produces a (nearly) full rank preliminary covariance matrix, which provides a relatively informative basis for constructing an accurate corrected covariance estimate. Finally, even at $\mathfrak{N}=80$, EnKF-MLC can still reduce $\bar\epsilon$ from $4.1$ to $1.5$, for which the drop is about $63\%$. All the above results for both Lorenz-63 and Lorenz-96 consistently prove the proposed EnKF-MLC framework’s capability to mitigate the filter divergence issue under various limited ensemble conditions.

\subsection{Sensitivity to Available Observations}
\label{sec:Observations}

Next, we evaluate the impact of the number of available observations on the performance of the proposed EnKF-MLC framework. For Lorenz-63, four observation configurations are tested, including $\{x,~y,~z\}$, $\{x,~y\}$, $\{x,~z\}$, and $\{y\}$. All other settings remain the same as the benchmark case and the results of $\bar\epsilon$ are shown in Fig.~\ref{fig:L63_rmse}(b) for both the traditional EnKF and EnKF-MLC methods. It is obvious that the EnKF-MLC consistently achieves substantially lower errors across all observation configurations. The most significant improvement occurs when all three variables ($\{x,~y,~z\}$) are observed, where $\bar\epsilon$ drops from $11.2$ (traditional EnKF) to $3.3$ (EnKF-MLC). Even under the most challenging condition with only a single variable ($\{y\}$) observed, the EnKF-MLC still reduces the $\bar\epsilon$ from $11.2$ to $4.8$, corresponding to a reduction of $57\%$. This demonstrates the robustness of the proposed method, even under scenarios of extremely sparse observation coverage.


For Lorenz-96, we consider three observation patterns, including all 40 variables $\{x^{(i)}~|~i=1:1:40\}$, every other variable (total 20)$\{x^{(i)}~|~i=1:2:39\}$, and every fourth variable (total 10)$\{x^{(i)}~|~i=1:4:37\}$. The results of $\bar\epsilon$ for the Lorenz-96 system are shown in Fig.~\ref{fig:L96_rmse}(b). Clearly, the EnKF-MLC consistently achieves substantially lower errors than the traditional EnKF across all observation configurations, with reductions of about $50\%$ for all cases. Specifically, in the most sparse case (every fourth variable, $\{x^{(i)}~|~i=1:4:37\}$), the traditional EnKF yields $\bar\epsilon = 5.1$, while the EnKF-MLC reduces it to $2.8$. These results, which are consistent with those of Lorenz-63, confirm that the proposed framework can always maintain a clear advantage over the traditional EnKF regardless of the observation coverage. 
\subsection{Sensitivity to DA Frequency}
\label{sec:DA Frequency}

Finally, we investigate the influence of the DA interval $T_{\text{DA}}$. For Lorenz‑63, we test two values of $T_{\text{DA}}$, namely 0.08~\text{MTU} and 0.25~\text{MTU}. As shown in Fig.~\ref{fig:L63_rmse}(c), the proposed EnKF-MLC framework exhibits robust performance across both DA intervals, maintaining consistently much lower values of $\bar\epsilon$ as compared to the traditional EnKF. For Lorenz-96, three DA intervals are considered, including $T_{\text{DA}} = 0.05,~0.10,~\text{and}~0.20$ MTU. As shown in Fig.~\ref{fig:L96_rmse}(c), the proposed EnKF-MLC algorithm can always reduce the error by approximately $50\%$, irrespective of the observation frequency.

\section{Conclusion}
\label{sec:Conclusion}

In this study, we propose a novel EnKF-MLC framework to tackle the common trade-off issue between accuracy and computational efficiency in the practical applications of traditional EnKF. Rather than discarding the ensemble completely, our approach leverages a limited ensemble and a MLP function to predict a covariance correction term. This correction is then applied to refine the forecast statistics within the standard EnKF workflow. Numerical experiments based on the Lorenz-63 and Lorenz-96 systems demonstrate that this EnKF-MLC framework consistently improves estimation accuracy across diverse DA configurations, while maintaining computational efficiency. These results highlight the value of integrating machine learning as a complementary tool to enhance traditional DA methods, particularly in scenarios where computational resources are limited.

\section*{Acknowledgements}
This research is financially supported by the Science and Technology Development Fund of Macau S.A.R. (0048/2025/ITP1, 001/2024/SKL and 0002/2025/EQP), the National Natural Science Foundation of China (52301336), and the University of Macau (SRG2025-00004-FST).

\bibliographystyle{elsarticle-harv} 
\bibliography{bibliography}






\end{document}